# Possible origin of the absence of magnetic order in LiOsO$_3$ : Spin-orbit coupling controlled ground state

Y. Zhang,[1] J. J. Gong,[1] C. F. Li,[1] L. Lin,[1] Z. B. Yan,[1] Shuai Dong,[2] and J. –M. Liu[1,3]

[1]*Laboratory of Solid State Microstructures, Nanjing University, Nanjing 210093, China*

[2]*School of Physics, Southeast University, Nanjing 211189, China*

[3]*Institute for Advanced Materials, Hubei Normal University, Huangshi 435002, China*

[**Abstract**] LiOsO$_3$ is the first experimentally confirmed polar metal with ferroelectric-like distortion. One puzzling experimental fact is its paramagnetic state down to very low temperature with negligible magnetic moment, which is anomalous considering its 5$d^3$ electron configuration since other osmium oxides (e.g. NaOsO$_3$) with 5$d^3$ Os ions are magnetic. Here the magnetic and electronic properties of LiOsO$_3$ are re-investigated carefully using the first-principles density functional theory. Our calculations reveal that the magnetic state of LiOsO$_3$ can be completely suppressed by the spin-orbit coupling. The subtle balance between significant spin-orbit coupling and weak Hubbard *U* of 5*d* electrons can explain both the nonmagnetic LiOsO$_3$ and magnetic NaOsO$_3$. Our work provides a reasonable understanding of the long-standing puzzle of magnetism in some osmium oxides.

# I. Introduction

Correlated electron systems with appreciable Coulomb repulsion are one of the most attractive platforms for accessing a series of emerging physical properties such as metal-insulator transition (MIT), superconductivity, colossal magnetoresistance, multiferroicity, and so on, which are often technologically useful. Such a Coulomb repulsion is usually characterized by the on-site Hubbard $U$. On the other hand, spin-orbit coupling (SOC) in condensed matters is becoming highly concerned, evidenced within a lot of emergent quantum materials such as topological insulators, Weyl semi-metals, and Kitaev systems [1-4], where SOC can be the core ingredient of physics underlying their novel physical phenomena. On one hand, SOC can be strong for heavy ions, and even comparable to $U$ for $5d$ electrons. For example, SOC is believed to play a decisive role in determining the unconventional properties in those $5d^5$ transition metal oxides [5], such as the $J_{\text{eff}}$=1/2 Mott state for iridates ($Ir^{4+}$) [6]. On the other hand, the wavefunctions of $5d$ electrons are more extended than those of $3d$ and $4d$ electrons, which effectively reduces the on-site Coulomb repulsion $U$. Therefore, the competitive and/or cooperative effect of SOC plus Coulomb repulsion provide a unique playground for novel $5d$ electronic properties.

Here we consider a specific $5d$ perovskite system: $LiOsO_3$, which is known as the first experimentally confirmed polar metal with ferroelectric-like structural transition [7], as predicted by Anderson and Blount [8]. Certainly, the major concern with such an unusual ferroelectric-like metallic state is not only the potential functionality but more importantly possible competition and coupling between the ferroelectricity and metallicity which are usually mutually exclusive.

A well-known but yet unsolved puzzling issue of $LiOsO_3$ is its magnetic ground state. In $LiOsO_3$, each Os ion is surrounded by an oxygen octahedron, which splits Os's $5d$ orbitals into the $t_{2g}$ and $e_g$ sectors by the crystalline field. In the ideal limit, the three $5d$ electrons of $Os^{5+}$ ion will occupy the $t_{2g}$ orbitals in the half filling manner. If the SOC effect is negligible, the half-filled $t_{2g}$ orbitals will lead to a total spin angular moment $S$=3/2 and a total orbital angular moments $L$=0, driven by the Hund's rule. Thus, the ideal magnetic moment should be 3 $\mu_B$/Os. Indeed, experiments show that most osmium oxides, e.g. $NaOsO_3$, $Cd_2Os_2O_7$, and $Ba_2YOsO_7$, have magnetic ground states, and their Curie/Nèel temperatures are ~69-410 K

[9-11]. However, although LiOsO$_3$ has a simple perovskite crystal structure and 5$d^3$ electron configuration similar to NaOsO$_3$, LiOsO$_3$ was experimentally found to show no any magnetic ordering even down to low temperature (~2 K) [7]. Later, a muon-spin relaxation ($\mu$SR) experiments also revealed the absence of magnetic order in LiOsO$_3$ down to 0.08 K [12].

There are several possible theoretical explanation for the absence of magnetic state in LiOsO$_3$. However, these theoretical proposals are rather confusing and often contradicting. For example, in Ref. [13], the local density approximate (LDA)+dynamical mean field theory (DMFT) calculations give a nonmagnetic (NM) to G-type antiferromagnetic (G-AFM) transition at ($U$=1.25 eV, $J_H$=0.1875 eV) or ($U$=0.7 eV, $J_H$=0.21 eV), while the LSDA+$U$ calculations in Ref. [14] report that the stable ground state should be a slater-type G-AFM insulator. In Ref. [15], ($U$=2.3 eV, $J_H$=0.345 eV) are adopted in the LDA+DMFT calculation which can lead to a large local moment (~2.5 $\mu_B$/Os) [13]. Then a crucial question is what are the proper values of $U$ (and $J_H$) for LiOsO$_3$, which are decisive for the ground state in the first-principles calculations. The resonant inelastic X-ray scattering (RIXS) experimental suggestion for $J_H$ is about 0.3 eV for osmium oxides [10, 16-17]. If so, the ground state should be magnetic in both LSDA+$U$ and LDA+DMFT calculations. Then the paramagnetism down to extreme low temperature remains a puzzle. Although the quantum fluctuation is expected to suppress the magnetic order and lead to so-called spin liquid state in some materials, here the large spin number $S$=3/2 seems to be not a proper candidate for the strong quantum fluctuation.

In this work, we will carefully re-investigate the electronic structure and magnetic ground state of LiOsO$_3$ as well as NaOsO$_3$ based on the density functional theory (DFT) calculations. The combined effect of SOC plus $U$ allows a comprehensive identification of the role of SOC, which has been somehow ignored in earlier studies.

## II. Computation methodology

The DFT calculations are performed using the pseudo-potential plane wave method as implemented in Vienna *ab initio* Simulation Package (VASP) [18-20]. The electron interactions are described using the Perdew-Burke-Ernzerhof (PBE) of the generalized gradient approximation (GGA) [21]. The projected augmented wave (PAW) [22] pseudo-potentials with a

500 eV plane-wave cutoff are used, including three valence electrons for Li ($1s^22s^1$), nine for Na ($2s^22p^63s^1$), fourteen for Os ($5p^66s^25d^6$), and six for O ($2s^22p^4$).

To investigate the combined effect of on-site Coulomb potential $U$ and SOC, we perform the GGA+$U$ and GGA+$U$+SOC calculations in details on a set of assigned magnetic structures, so that the interplay of $U$ and SOC can be clarified.

The low temperature structures of LiOsO$_3$ and NaOsO$_3$ are shown in Fig. 1(a) and 1(b) respectively. Starting from the experimental structures, the lattice constants and all atomic coordinates are fully relaxed within the initial space groups, until the Hellman-Feynman forces on every atom are converged to less than 1.0 meV/Å. To accommodate the magnetic structure, $2 \times 2 \times 2$ supercell of LiOsO$_3$ containing 80 atoms to build various types of antiferromagnetic (AFM) orders. A $20 \times 20 \times 20$ mesh for the unit cell of LiOsO$_3$, a $7 \times 7 \times 7$ mesh for the supercell of LiOsO$_3$, and a $13 \times 13 \times 13$ mesh for the unit cell of NaOsO$_3$, are used for the Brillouin-zone sampling.

Here four magnetic ordered structures: A-type AFM (A-AFM) order, C-type AFM (C-AFM) order, G-AFM order, ferromagnetic (FM) order, are considered in the present calculations, plus the NM state. The three AFM orders are sketched in Fig. 2(a-c).

## III. Results and discussion

To solve the aforementioned confusing theoretical results, it is necessary to clarify the methods of +$U$ in the first-principles calculations. Taking the most used VASP code for example, there are three choices of +$U$: a) LDAUTYPE=1: the rotationally invariant LSDA+$U$ introduced by Liechtenstein et al. [23]; b) LDAUTYPE=2: the simplified (rotationally invariant) approach to the LSDA+$U$, introduced by Dudarev et al. [24]; c) LDAUTYPE=4: LDA+$U$. The LDAUTYPE=2 is the most used (default) choice, which only needs a parameter $U_{eff}$=$U$-$J$. For the LSDA calculation (LDAUTYPE=1 and 2), the exchange splitting, e.g. the effect of $J_H$, has already (partially) included, even without $U$. In other words, the $U$ and $J_H$ in these two choices are not the naked ones as used in the Hubbard models, or DMFT calculations, or RIXS experiments, but significantly reduced. Instead, the rarely-used LDA+$U$ (LDAUTYPE=4) choice can give naked $U$ and $J_H$ to compare between the DFT, model, DMFT, as well as RIXS experiment.

First, the LDA+$U$ (LDAUTYPE=4) choice is tested for LiOsO$_3$ and the results are shown in Fig. 3. Two cases of $J/U$ (=0.15 & 0.3) are considered for example. The lattice constant $a$ is a little larger than the experimental one, and gradually increases with $U$ (Fig. 3(a)). It is well known that GGA will systematically and slightly overestimated the lattice constants. Despite this point, our result agrees with the experimental value, especially in the low $U$ region. The G-AFM state is the lowest energy one among all magnetic candidates. Thus, the energy difference between G-AFM and NM state is shown Fig. 3(b), which becomes negligible in the low $U$ region, e.g. $U \leqslant 0.9$ eV for $J/U$=0.3 or $U \leqslant 1.3$ eV for $J/U$=0.15. Not surprisingly, the energy degeneration in the low $U$ region is due to the quenching of local moment of Os, although the critical value of $U$'s for zero local moment is a little bit lower for 0.2-0.3 eV, as shown in Fig. 3(c). With increasing $U$, the metal-insulator transition occurs almost accompanying the NM-G-AFM transition (Fig. 3(d)), with slightly shift of critical $U$ for 0.1-0.2 eV higher. These results obtained in our LDA+$U$ calculation agrees with previous LDA+DMFT results.

By calculating more points of $J$ and $U$, a phase diagram can be sketched as Fig. 4. In addition, the calculation with SOC is also performed, which can slightly shift the NM-G-AFM boundary to larger $U$ and $J$ side. As expected, the NM state exists in the low $U$ and low $J$ region. Taking the experimental value of $J_H$~0.3 eV for reference, the ground state is probably located at the boundary between NM and G-AFM state, with zero or very small local moment (<0.25 $\mu_B$/Os), instead of large local moment (~2.5 $\mu_B$/Os) obtained in Ref. [15]. Thus it is probably that LiOsO$_3$ is indeed NM with almost zero or very small local moment, which can properly understand the paramagnetism down to very low temperatures.

Then it is interesting to know whether the more commonly used LSDA+$U$ method can correctly describe the nonmagnetism/magnetism of LiOsO$_3$. By setting LDAUTYPE=2, the same processes have been done, whose results are summarized in Fig. 5. Our LSDA+$U$ results lead to magnetic ground state with large local moment (>1.1 $\mu_B$/Os), which further increases with $U_{\text{eff}}$. So the LSDA+$U$ calculation could not explain the experimental fact, then we consider the effect of SOC. By considering the SOC, the magnetic moment is reduced by residual orbital moment and the phase diagram is significantly changed. The NM state becomes the ground state in the low $U_{eff}$ region (≤0.3 eV). The transition from NM to G-AFM is the first order with discontinuous jump of local moment. Physically, the SOC coupling (expressed as $\boldsymbol{L}*\boldsymbol{S} = L^zS^z + (L^+S^-+L^-S^+)/2$, where $\boldsymbol{L}$ and $\boldsymbol{S}$ are orbital and spin operators) can mix the spin-up and spin-down channels due to the raising and lowering operators $L^+$ and $L^-$, which would reduce the effective Hubbard repulsion between spin-up and spin-down channels. In short, the SOC plays nonnegligible role to obtain the NM ground state of $LiOsO_3$, at least in the LSDA+$U$ calculation. Thus, the mostly used LSDA+$U$+SOC method can also describe the magnetic fact of $LiOsO_3$.

Then it is necessary to check this method in $NaOsO_3$, since a successful theoretical approach should be valid for various systems, at least for a family of materials. With the same LSDA+$U$ method, our calculation confirms that the ground state for $NaOsO_3$ is G-AFM state, which is robust against the SOC, as summarized in Fig. 6. This result is different from $LiOsO_3$, but agrees with the experimental factor, further confirming the LSDA+$U$ (+SOC) method can describe these osmium oxides. For the moment of $Os^{5+}$ in the G-AFM $NaOsO_3$ as a function of $U_{eff}$, the calculated value at $U_{eff}$=0 eV is 1.01 $\mu_B$ and 1.38 $\mu_B$ at $U_{eff}$=1.0 eV. It is noted that the measured moment for $NaOsO_3$ is about 1.0 $\mu_B$, obtained from neutron scattering [25]. This implies convincingly $U_{eff}$ ~ 0 eV in the LSDA+$U$ calculations for $NaOsO_3$. And the density of states (DOS) of $NaOsO_3$ in G-AFM state with $U_{eff}$ =0 is shown in Fig. 6(c), which is consistent with a pure Slater-type insulator, as confirmed in experiments [25].

To better understand the contrastive magnetism of $LiOsO_3$ and $NaOsO_3$, the structural differences are shown in Table I. It can be seen that the Os-O-Os network is more compact in $LiOsO_3$, with averagely shorter Os-O bonds and short distance between nearest-neighbor Os-Os. Thus, the hybridization between Os's $5d$ and O's $2p$ orbitals are more prominent in $LiOsO_3$ and the effective hopping between nearest-neighbor Os's $5d$ orbitals are larger. To further confirm this point, the partial DOS's of each Os in $LiOsO_3$ and $NaOsO_3$ in the G-AFM state with $U_{eff}$=0, are shown in Fig. 7. As expected, the band width of $5d$ orbitals in $LiOsO_3$ is relative wider than that in $NaOsO_3$. It is well known that the Mott transition (also the magnetic transition) of Hubbard model system depends on the subtle competition between the kinetic energy and Coulombic repulsion. The cases of $Os^{5+}$ are just located around the critical point of Mott transition. The narrower $5d$ bands of $NaOsO_3$ is advantaged for magnetic moment, while the wider 5$d$ bands of LiOsO$_3$ prefers the nonmagnetic metallic state.

## IV. Conclusion

The magnetic and electronic properties of LiOsO$_3$ and NaOsO$_3$ have been checked using first-principles methods. The long-standing puzzle regarding the paramagnetism of LiOsO$_3$ has been clarified. In our opinion, the local magnetic moment of Os$^{5+}$ in LiOsO$_3$ can be zero or very small, due to the weak Hubbard $U$ and indispensable SOC. The LSDA+$U$+SOC method can describe the magnetism of LiOsO$_3$, although the $U$ used in LSDA is significantly reduced comparing with the value used in LDA+$U$ or LDA+DMFT. In contrast, the magnetic ground state has been verified for NaOsO$_3$. The importance of SOC in determining the magnetic structure has already been proved in some other 5$d^3$ osmium oxides such as Sr$_2$ScOsO$_6$ and Cd$_2$Os$_2$O$_7$ [16, 26]. Our work provides a uniform description for LiOsO$_3$ and other osmium oxides.

## ACKNOWLEDGMENTS

This work was supported by the National Key Research Program of China (Grant Nos. 2016YFA0300101 and 2015CB654602), the Natural Science Foundation of China (Grant Nos. 51431006 and 51332006).

**Table Caption:**

Table I. Structural parameters of LiOsO$_3$ and NaOsO$_3$ calculated using LSDA with G-AFM order.

|  | LiOsO$_3$ | NaOsO$_3$ |
|---|---|---|
| Os-O bond length (average) | 1.943 Å | 1.951 Å |
| Nearest-neighbor Os-Os distance | 3.654 Å | 3.788 Å |

**Figure Captions:**

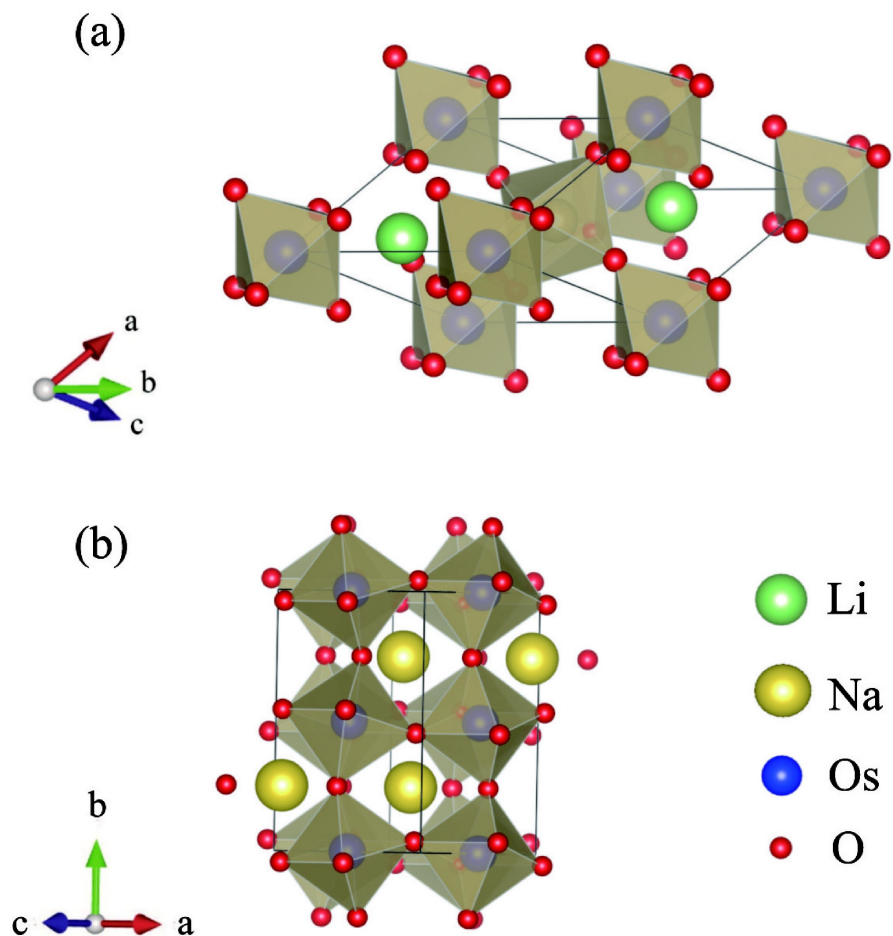

Figure 1. Crystal structure of (a) LiOsO$_3$ and (b) NaOsO$_3$. The green, yellow, blue, and red balls are Li, Na, Os, and O ions.

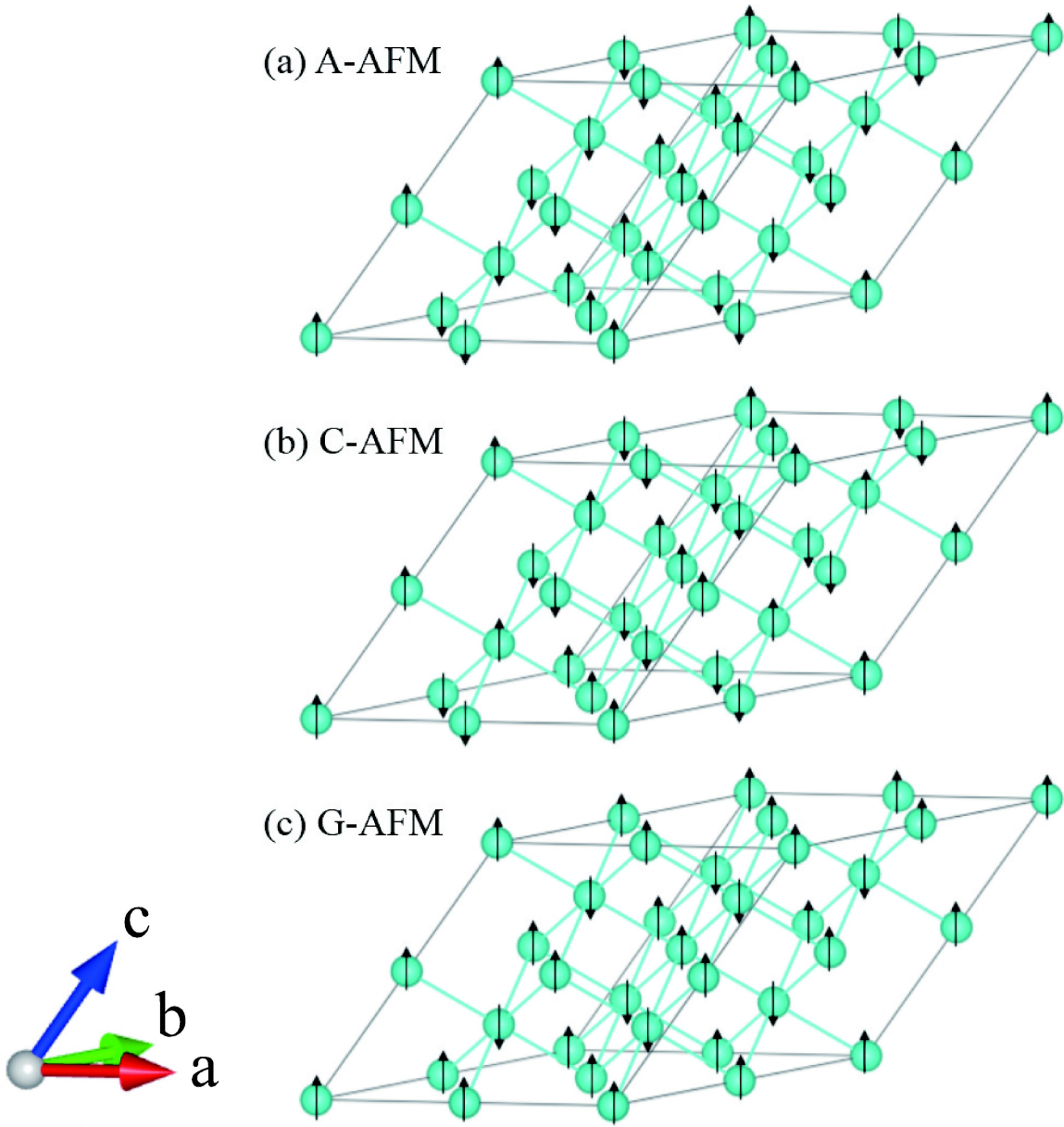

Figure 2. Schematic of three candidate antiferromagnetic orders for LiOsO$_3$. (a) A-AFM, (b) C-AFM, (c) G-AFM in the 2×2×2 supercell. Only Os ions are shown and the arrows represent the signs of magnetic moments.

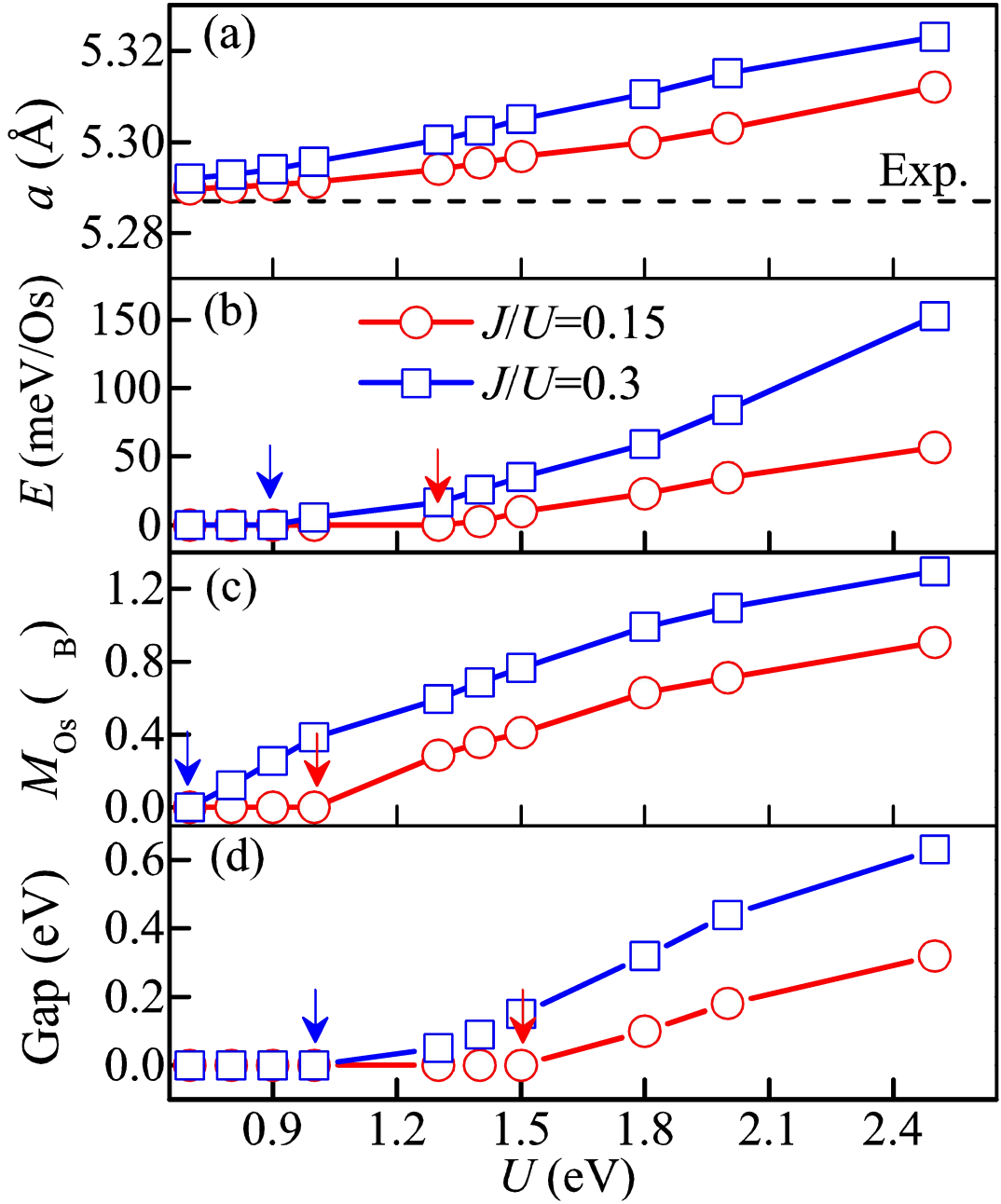

Figure 3. Results of LDA+$U$ calculations for LiOsO$_3$ as a function of $U$. Two ratios of $J/U$ are considered. (a) The optimized lattice constant in G-AFM state. (b) The energy difference between G-AFM state and NM state ($\Delta E = E_{\text{G-AFM}} - E_{\text{NM}}$) (c) The magnetic moment of Os ion. (d) The energy gap of LiOsO$_3$ in G-AFM state.

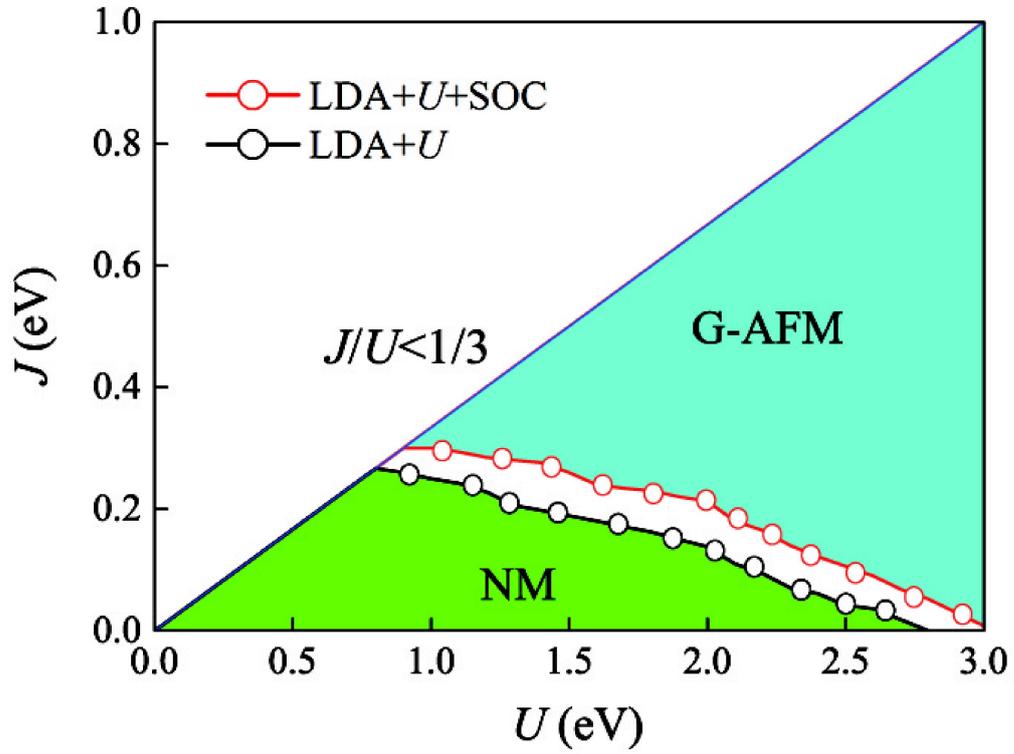

Figure 4. The magnetic phase diagram of LiOsO$_3$ calculated by LDA+$U$ and LDA+$U$+SOC. Physically, $J/U$ should be less than 1/3. The phase boundary between NM and G-AFM is shifted to higher $U$ and $J$ side by SOC.

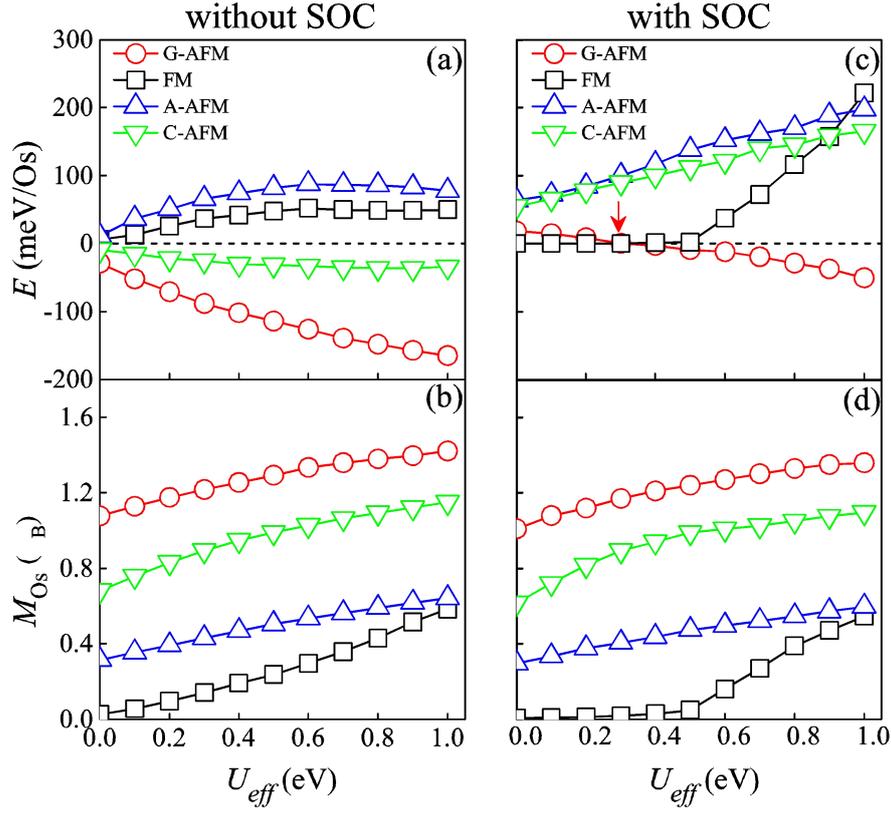

Figure 5. Results of LSDA+$U$ and LSDA+$U$+SOC calculations for LiOsO$_3$ as a function of $U_{eff}$. (a-b) Without SOC. (c-d) With SOC. (a) and (c) The energy difference $\Delta E$ between the magnetic phases and NM one. (b) and (d) The local Os$^{5+}$ moment.

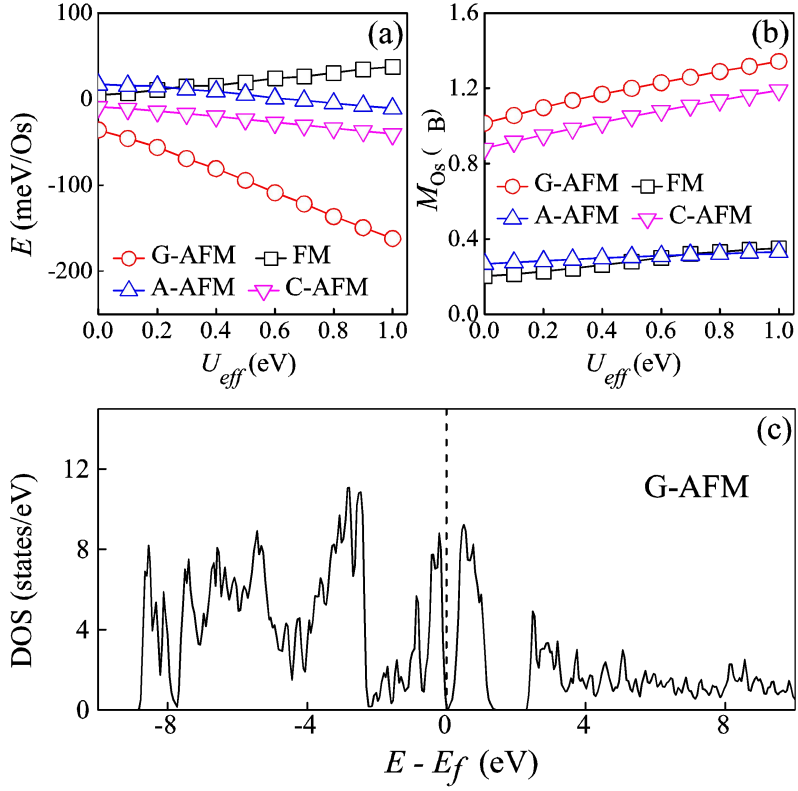

Figure 6. Results of LSDA+$U$+SOC calculations for NaOsO$_3$ as a function of $U_{\text{eff}}$. (a) The energy difference $\Delta E$ between the magnetic phases and NM one. The G-AFM order is always the lowest one. (b) The local Os$^{5+}$ moment, which is always nonzero. (c) The total density of states (DOS) of NaOsO$_3$ when $U_{\text{eff}} = 0$.

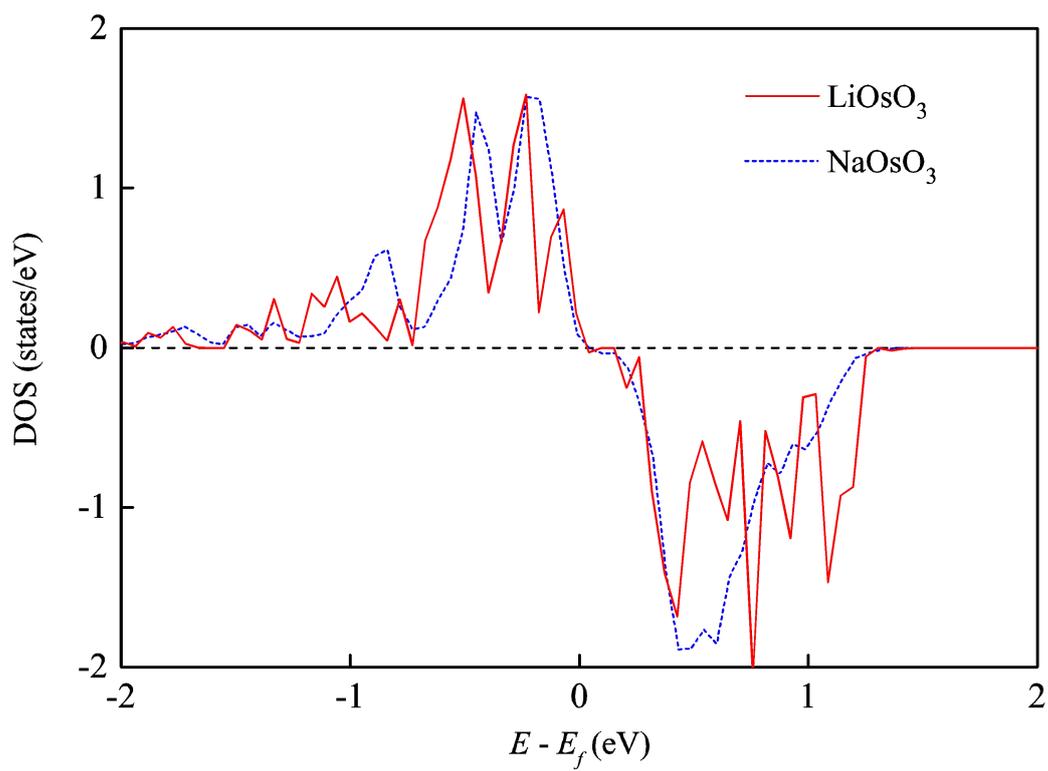

Figure 7. The partial spin polarized DOS of a single spin-up Os in LiOsO$_3$ and NaOsO$_3$ in the G-AFM state calculated by LSDA. Here only the differences of spin-up and spin-down channel are shown.